# New example of CP violation from search for the permanent electric dipole moment of Cs atoms

Pei-Lin You [1]    Xiang-You Huang [2]
1. Institute of Quantum Electronics, Guangdong Ocean University, zhanjiang 524025, China.
2. Department of Physics, Peking University, Beijing 100871, China.

Using special capacitors three experiments to search for a permanent electric dipole moment (EDM) of Cesium atom were completed. The electric susceptibility($x_e$) of Cs vapor varies in direct proportion to the density N, where $x_e \geq 70$ when $N \geq 7.37 \times 10^{22}$ m$^{-3}$! The relationship between $x_e$ of Cs vapor and the absolute temperatures T is $x_e = B/T$, where the slope $B \approx 320(k)$ as polar molecules $H_2O$ ($B \approx 1.50(k)$). Its capacitance C at different voltage V was measured. The C-V curve shows that the saturation polarization of Cs vapor has be observed when the field $E \geq 7.4 \times 10^4$ V/m. Our measurements give the EDM of an Cs atom : $d_{Cs} = [2.97 \pm 0.36(\text{stat}) \pm 0.24(\text{syst})] \times 10^{-29}$ C.m. New example of CP (charge conjugation and parity) violation occurred in Cs atoms. Our results are easy to be repeated because the details of the experiment are described in the article.

**1. Introduction**    In order for an atom or elementary particle to possess a permanent electric dipole moment (EDM), time reversal (T) symmetry must be violated, and through the CPT theorem CP(charge conjugation and parity) must be violated as well[1]. The currently accepted Standard Model of Particle Physics predicts unobservable the dipole moments of an atom, therefore, EDM experiments are an ideal probe for new physics beyond the Standard Model. Experimental searches for EDMs can be divided into three categories: search for the neutron EDM[2], search for the electron EDM utilizing paramagnetic atoms, the most sensitive of which is done with Tl atoms(the result is $d_e = [1.8 \pm 1.2(\text{statistical}) \pm 1.0(\text{systematic})] \times 10^{-27}$ e.cm)[3], and search for an EDM of diamagnetic atoms, the most sensitive of which is done with $^{199}$Hg(the result is $d(Hg) = -[1.06 \pm 0.49(\text{stat}) \pm 0.40(\text{syst})] \times 10^{-28}$ e.cm )[1,4]. Experiments to search for an EDM of atom began many decades ago, no large EDM has yet been found [1-5]. In all experiments, they measured microcosmic Larmor precession frequency of individual particle based on nuclear spin or electron spin. The search for an EDM consists of measuring the precession frequency of the particle in parallel electric and magnetic fields and looking for a change of this frequency when the direction of **E** is reversed relative to **B.** We now submit the article on the similar topic, however, with measuring macroscopic electric susceptibility($x_e$) of Cs vapor containing a large number of Cs atoms (the density $N > 10^{21}$ m$^{-3}$). This article reported three new experimental phenomenon which have not been observed during the past few decades. Our experiments showed that ground state Cs atom may have a large EDM. On the other hand, some evidence for CP violation beyond the Standard Model comes from Cosmology. Astronomical observations indicate that our Universe is mostly made of matter and contains almost no anti-matter. The first example of CP violation was discovered in 1964, but it has been observed only in the decays of the $K_o$ mesons. After 38 years, the BaBar experiment at Stanford Linear Accelerator Center (SLAC) and the Belle collaboration at the KEK laboratory in Japan announced the second example of CP violation. "The results gave clear evidence for CP violation in B mesons. However, the degree of CP violation now confirmed is not enough on its own to account for the matter-antimatter imbalance in the Universe." (SLAC Press Release July 23, 2002).This fact suggests that there must be other ways in which CP symmetry breaks down and rarer processes and more subtle effects must be examined. So EDM experiments are now considered an ideal probe for evidence of new sources of CP violation. If an EDM is found, it will be compelling evidence for the existence of new sources of CP violation. Our experiments showed that new example of CP violation occurred in Cs atoms. **The results could offer the first hints of a breakdown of the Standard Model. This finding is a vital clue that an unknown factor is very likely at play in Cs atoms**[6]**.** The correctness of our theoretical motivation is supported by such five facts as the following.

①The shift in the energy levels of an atom in an electric field is known as the Stark effect. Normally the effect is quadratic in the field strength, but first excited state of the hydrogen atom exhibits an effect that is linear in the strength. This is due to the degeneracy of the excited state. This result shows that the hydrogen atom (the quantum number n=2 ) has very large EDM, $d_H = 3ea_o = 1.59 \times 10^{-8}$ e.cm ($a_o$ is Bohr radius)[7,8]. On the one hand, this EDM does not depend on the field strength, hence it is not induce by the external field but is inherent behavior of the atom[7,8]. L.I. Schiff once stated that "Unperturbed degenerate states of opposite parities, as in the case of the hydrogen atom, can give rise to a permanent electric dipole moment"[7]. L.D. Landay also once stated that "The presence of the linear effect means that, in the unperturbed state, the hydrogen atom has a dipole moment"[8]. On the other hand, the calculation of quantum mechanics tells us that unperturbed degenerate states of hydrogen atom(n=2) with zero EDM and has result $\langle \psi_{2lm} | er | \psi_{2lm} \rangle = 0$, where $\psi_{2lm}$ are four wave functions of unperturbed degenerate states[8]. Due to the EDM of the hydrogen atom is responsible for the presence of linear Stark effect. A hydrogen atom(n=2) with zero EDM how responds to the external field and results in the linear Stark effect ? Quantum mechanics can not answer the problem [9-11]!

②In addition, the radius of the hydrogen atom of the first excited state is $r_H = 4a_o = 2.12 \times 10^{-8}$ cm, it is almost the same as the radius of $^{199}$Hg ($r_{Hg} = 1.51 \times 10^{-8}$ cm)[12], but the discrepancy between their EDM is by some twenty orders of magnitude! How do explain this inconceivable discrepancy? The existing theory can not answer the problem! The existing theory thinks that in quantum mechanics there is no such concept as the path of an electron[8]. No one will give you any deeper explain of the inconceivable discrepancy. No one has found any more basic mechanism from which these results can be deduced! However, a hydrogen atom (n=2) has a nonzero EDM in the semi-classical theory of atom. The electron in a hydrogen atom (n=2) moves along a quantization elliptic orbit. We can draw a straight line perpendicular to the major axis of the elliptic orbit through the nucleus in the orbital plane. The straight line divides the elliptic orbit into two parts that are different in size. According the law of conservation of angular momentum, the average distance between the moving electron and the static nucleus is larger and the electron remains in the large part longer than in the small part. As a result, the time-averaged value of the electric dipole moment over a period is nonzero.

③The alkali atoms having only one valence electron in the outermost shell can be described as hydrogen-like atoms[13]. In the Sommerfeld motel the valence electron moves along a highly elliptical orbit, the so-called diving orbits, approach the nucleus closely. Furthermore, since the quantum number of the ground state alkali atoms are n≥2 rather than n=1( this is 2 for Li, 3 for Na, 4 for K, 5 for Rb and 6 for Cs), as the excited state of the hydrogen atom. So we conjecture that the ground state neutral alkali atoms, as the excited state of the hydrogen atom, may have large EDM of the order of e $a_o$, but the actual result of the conjecture still needs to be tested by experiments[14].

④When atoms are placed in an electric field, they become polarized, acquiring *induced* electric dipole moments in the direction of the field. On the other hand, many molecules do have EDM. This molecule is called a polar molecule, such as $H_2O$, HCl etc. When an electric field is applied, the polar molecules tend to orient in the direction of the field. Note that the susceptibility($x_e$) caused by the orientation of polar molecules is inversely proportional to the absolute temperature(T): $x_e = B/T$ while the induced susceptibility due to the distortion of electronic motion in atoms is temperature independent: $x_e = A$, where A and the slope B is constant, $x_e = C/C_o - 1$, $C_o$ is the vacuum capacitance and C is the capacitance of the capacitor filled with the material[15]. J.D. Jackson once stated that this difference in temperature dependence offers a means of separating the polar and non-polar substances experimentally[15]. The molecular electric dipole moments are of the order of magnitude of the electronic charge multiplied by the molecular dimensions, or about $10^{-30}$ coulomb-meters.

⑤R.P. Feynman considered the orientation polarization of water vapor molecules[16]. He plotted the straight line from four experimental points and $x_e = A + B/T \approx B/T$, where $A \approx 0$. $x_e$ is given by the expression

$$x_e = N d_o^2 / 3kT \varepsilon_o \qquad (1)$$

where k is Boltzmann constant, $\varepsilon_o$ is the permittivity of free space, N is the number density of gas, $d_o$ is the EDM of a molecule and the slope $B = N d_o^2 / 3k \varepsilon_o$. Table 1 gives the four experimental data[17].

**Table 1 Experimental measurements of the susceptibility of water vapor at various temperatures**

| T(K) | Pressure(cm Hg) | $x_e \times 10^5$ |
| --- | --- | --- |
| 393 | 56.49 | 400.2 |
| 423 | 60.93 | 371.7 |
| 453 | 65.34 | 348.8 |
| 483 | 69.75 | 328.7 |

From the ideal gas law, the average density of water vapor is $N = P/kT = (1.392 \pm 0.002) \times 10^{25}$ m$^{-3}$. We work out the slope B =(1.50 ±0.04)K and the EDM of a water molecule is $d_{H2O} = (3\varepsilon_o B/N)^{1/2} = (6.28 \pm 0.18) \times 10^{-30}$ C. m. The result in agreement with the observed value $d_{H2O} = 6.20 \times 10^{-30}$ C. m[18]. In addition, from $B \approx 0.94$ K the EDM of a HCl molecule can be deduced to be $d_{HCl} = 3.60 \times 10^{-30}$ C. m $\approx 0.43 ea_o$ [18]. **If Cs atom is** the polar atom, a temperature dependence of the form $x_e = B/T$ should be expected in measuring $x_e$ at different temperatures and we may calculate the EDM of an Cs atom using the same method!

**2. Experimental method and result** The first experiment: investigation of the relationship between the capacitance C' of Cs or Hg vapor and the number density N of atoms. The experimental apparatus were two closed glass containers containing Cs or Hg vapor. Two silver layers **a** and **b** build up the glass cylindrical capacitor(Fig.1,where the plate area $S_1=(5.8\pm0.1)\times10^{-2}$ m$^2$, the plate separation $H_1=(9.6\pm0.1)$mm). The radiuses of the layers **a** and **b** are shown respectively by r and R. Since R-r=H+2△<< r (△=1.2mm), the capacitor can be approximately regarded as a parallel-plane capacitor. The capacitance was measured by a digital meter. The precision of the meter was 0.1pF, the accuracy was 0.5% and the surveying voltage was V=(1.0±0.01)volt. It means that the applied field $E=V/H_1\approx1.1\times10^2$v/m is weak using the meter. This capacitor is equivalently connected in series by two capacitors. One is called C' and contains the Cs or Hg vapor of thickness H , another one is called C'' and contains the glass medium of thickness 2△. The total capacitance C can be written as C=C'C''/ (C'+C'') or C'=CC''/ (C'' - C), where C'' and C can be directly measured. When the two containers are empty, they are pumped to a vacuum pressure P≤10$^{-6}$ Pa for 20 hours. The measured capacitances are nearly the same: C'$_{10}$=(54.0±0.1)pF(for Cs) and C'$_{20}$=(52.8±0.1)pF(for Hg). Next, a small amount of the Cs or Hg material with high purity is put in the two containers respectively in a vacuum environment. The two containers are again pumped to vacuum pressure P ≤10$^{-6}$ Pa , then they are sealed. Now, their capacitances are C'$_1$=(102.2±0.1)pF and C'$_2$=(53.6±0.1)pF respectively at room temperature. We put the two capacitors into a temperature-control stove, raise the temperature of the stove very slow and keep at $T_1$=(473±0.5)K for 3 hours. It means that the readings of capacitance are obtained under the condition of Cs or Hg saturated vapor pressure. Two comparable experimental curves are shown in Fig.2. From Ref.[12], the saturated pressure of Hg vapor $P_{Hg}$=2304.4 Pa at 473K and measuring capacitance C'$_{2t}$= (56.4±0.1) pF. From the ideal gas law, the density of Hg vapor is $N_{Hg} = P_{Hg}/kT_1 = 3.53\times10^{23}$ m$^{-3}$. The formula of saturated vapor pressure of Cs vapor is **P=10$^{6.949-3833.7/T}$** psi and the effective range is 473K ≤T ≤623K[12]. We obtain the saturated pressure of Cs vapor $P_{Cs}$=0.0698 psi =481.3Pa at 473K and measuring capacitance C'$_{1t}$= (3944±10) pF. The density of Cs atoms is $N_{Cs} = N_1 = P_{Cs}/kT_1 = 7.37\times10^{22}$ m$^{-3}$. The experiment shows that the number density $N_{Hg}$ of Hg vapor is 4.79 times as $N_{Cs}$ of Cs vapor but the capacitance change(C'$_{2t}$ - C'$_{20}$) of Hg vapor being only 1/1080 of (C'$_{1t}$ - C'$_{10}$) of Cs vapor! So unlike Cs atom, a Hg atom is non-polar atom.

The second experiment: investigation of the relationship between the electric susceptibility($x_e$) of Cs or Hg vapor and temperatures(T) under each fixed density. The experimental apparatus was a closed glass container containing Cs vapor. Two stainless steel tubes **a** and **b** build up the glass cylindrical capacitor (Fig.3). Since R-r=H<<r, the capacitor could be approximately regarded as a parallel-plane capacitor. The capacitance was still measured by the digital meter, and vacuum capacitance $C_0$=(66.0±0.1)pF(where $S_2$= (5.7±0.1)×10$^{-2}$ m$^2$, $H_2$=(7.6±0.1)mm ). Then the capacitor filled with Cs vapor at the fixed density $N_2$ and was put into the stove. By measuring the electric susceptibility $x_e$ of Cs vapor at different temperature T, we obtain $x_e$=A+B/T ≈B/T, where the slope B=(320±4) K. The intercept A represents the susceptibility due to induced polarization and A≈0.The capacitance of an identical capacitor containing Hg vapor was measured and we found that the slope is nearly zero, B≈0.0K( $x_e$<0.07 is nearly constant). The experimental results are shown in Fig.4.

The third experiment: measuring the capacitance of Cs vapor at various voltages (V) under a fixed $N_2$ and $T_2$. The apparatus was the same as the second experiment ($C_0$=(66.0±0.1)pF) and the method is shown in Fig.5. C was the capacitor filled with Cs vapor to be measured, which was kept at $T_2$= (323±0.5)K, and $C_d$ was used as a standard capacitor. Two signals $V_c(t)=V_{co} \cos\omega t$ and $V_s(t)=V_{so} \cos\omega t$ were measured by a digital oscilloscope having two lines. The two signals had the same frequency and always the same phase at different voltages when the frequency is higher than a certain value. It indicates that capacitor C filled with Cs vapor was the same as $C_d$, a pure capacitor without loss. From Fig.5, we have C = $C_d$ ($V_{so}$–$V_{co}$)/$V_{co}$. In the experiment $V_{so}$ can be adjusted from zero to 800V. The capacitance C at various voltages was shown in Fig.6. When $V_{co}=V_1$=(0.3±0.002)volt, $C_1$=(130.0±0.2)pF is approximately constant, and $x_e=C_1/C_o$ - 1=0.9697. With the increase of voltage, the capacitance decreases gradually. When $V_{co}=V_2$=(560±2)V, $C_2$ =(68.0±0.2) pF, it approaches saturation and $x_e =C_2/C_0$ - 1≤0.0303≈0. If nearly all dipoles in a gas turn toward the direction of the field, or $x_e$ of the gas is approximately the same as the vacuum, this effect is called the saturation polarization. The C-V curve shows that the saturation polarization of Cs vapor is obvious when E= $V_2/H_2 \geq 7.4\times10^4$v/m.

**3. Discussion**

①Our experimental apparatus is a closed glass container. It resembles a Dewar bottle flask in shape. In all experiments, when the capacitor is empty, it is pumped to vacuum pressure P ≤$10^{-6}$ Pa for 20 hours. The aim of the operation is to remove carefully impurities, such as oxygen, absorbed on the inner walls of the container. In addition, the purity of Cs material exceeded 0.9998. **Actual result showed that the Cs vapor filled the capacitors is present in atomic form, not the dimer.**

②The electric susceptibility of polar molecules is of the order of $10^{-3}$ for gases at NTP and $10^0$ (or one) for solids or liquid at room temperature [15]. Experimentally, typical values of the susceptibility are 0.0046 for HCl gas, 0.007 for water vapor, 5-10 for glass, 3-6 for mica[18]. When the field is weak using the digital meter, our experiment showed that the electric susceptibility of Cs vapor was measured to be **$x_e = C'_1/C'_{10} - 1 > 0.8$** (where the density N≈$10^{21}$m$^{-3}$), and **$x_e = C'_{1t}/C'_{10} - 1 ≈ 72 >> 1$** (where N=7.37×$10^{22}$ m$^{-3}$) ! **Few experiments in atomic physics have produced a result as surprising as this one.**

③The electric susceptibility caused by the orientation of gaseous polar molecules is[19]

$$x_e = N d_o L(a)/\varepsilon_o E \qquad (2)$$

The mean value of cos θ : $<\cos θ> = \mu \int_0^\pi \cos θ \exp(d_o E \cos θ /kT) \sin θ dθ = L(a)$, where a = $d_o$ E /kT , μ is a normalized constant, θ is the angle between **$d_o$** and **E**. L(a) = [($e^a + e^{-a}$) / ($e^a - e^{-a}$)] – 1/a is called the Langevin function, where a>>1, L(a)≈1; a<<1, L(a)≈a/3 and we get Eq.(1)[19]. When the saturation polarization appeared, L(a)≈1 and this will happen only if a>>1[19]. R.P. Feynman once stated that *" when a filed is applied, if all the dipoles in a gas were to line up, there would be a very large polarization, but that does not happen"* [16]. **So, no scientist has observed the saturation polarization effect of any gaseous dielectric till now!** The saturation polarization of Cs vapor in ordinary temperatures is an entirely unexpected discovery.

④Due to the induced dipole moment of Cs atom is $d_{int} = G \varepsilon_o E$, where G=59.6×$10^{-30}$m$^3$ [20], the most field strength is Emax≤7.4×$10^4$v/m in the experiment, then $d_{int}$≤3.9×$10^{-35}$ C.m can be neglected. We obtain

$$x_e = N d L(a)/\varepsilon_o E \qquad (3)$$

where d is the EDM of an Cs atom and N is the number density of Cs atoms. **L(a)= <cos θ> is the ratio of Cs atoms which lined up with the field in the total atoms.** Note that $x_e = \varepsilon_r - 1 = C/C_o - 1$, where $\varepsilon_r$ is the dielectric constant , E=V/H and $\varepsilon_o = C_o H / S$, leading to

$$C - C_o = \beta L(a)/a, \qquad (4)$$

**This is the polarization equation of Cs atoms**, where β = S N d$^2$/kTH is a constant. Due to a=d E/k T= dV/kTH **we obtain the formula of atomic EDM**

$$\mathbf{d_{atom} = (C - C_o)V / L(a)SN} \qquad (5)$$

In order to work out **L(a)** and **a** of the first experiment, note that in the third experiment when the field is weak ($V_1$=0.3 v), $a_1$<<1 and L($a_1$)≈$a_1$/3, $C_1 - C_o = \beta /3$ and β =192 pF. When the field is strong($V_2$= 560 v), $a_2$>>1 and L($a_2$)≈1, $C_2 - C_o = L(a_2) \beta /a_2 ≈ \beta /a_2$. We work out $a_2 ≈ \beta /(C_2 - C_o)$=96, L($a_2$)≈L(96)=0.9896, $a_2$ = β L($a_2$)/($C_2 - C_o$)=95, L($a_2$)=L(95)=0.9895. Due to **a**=d E/kT=dV/kTH, so **a/$a_2$** =V$T_2H_2$/$T_1H_1V_2$. Substituting the corresponding values, we obtain **a**=0.0917 and L(**a**)≈**a**/3=0.0306. L(**a**)≈0.0306 means that only 3.06% of Cs atoms have be oriented in the direction of the field when the field E=V/$H_1$=1.1×$10^2$V/m. Substituting the values: $S_1$=5.8×$10^{-2}$ m$^2$, $N_1$=7.37×$10^{22}$ m$^{-3}$, V=1.0volt ,C - $C_o$ = $C'_{1t} - C'_{10}$ = 3890pF and L(a), we work out

$$d_{Cs} = (C - C_o)V / L(a) S_1 N_1 = 2.97×10^{-29} C.m = 1.86×10^{-8} e.cm ≈ 3.5 e a_o \qquad (6)$$

The statistical errors is △$d_1$/d=△C/C+△$C_o$/$C_o$+△$S_1$/$S_1$+△V/V+△$N_1$/$N_1$＜0.12, considering all systematic error △$d_2$/d＜0.08 (including △L(a)/L(a)), and the combination error △d/d＜0.15. We find that

$$\mathbf{d(Cs)=[2.97±0.36(stat)±0.24(syst)]×10^{-29}C.m = [1.86±0.22(stat)±0.15(syst)]×10^{-8}e.cm} \qquad (7)$$

**Although above calculation is simple, but no physicist completed the calculation up to now!**

⑤The formula $d_{atom} = (C - C_o)V/L(a)SN$ can be justified easily. The magnitude of the dipole moment of an Cs atom is d = e r. N is the number of Cs atoms per unit volume. L(a) is the ratio of Cs atoms lined up with the field in the total number. Suppose that the plates of the capacitor have an area S and separated by a distance H, the volume of the capacitor is SH. When a field is applied, the Cs atoms tend to orient in the direction of the field. On the one hand, the change of the charge of the capacitor is △Q=(C –$C_0$)V. On the other hand, when the Cs atoms are polarized by the orientation, the total number of Cs atoms lined up with the field is SHNL(a).The number of layers of Cs atoms which lined up with the filed is H/r. Because inside the Cs vapor the positive and negative charges cancel out each other, the polarization only gives rise to a net positive charge on one side of the capacitor and a net negative charge on the opposite side. Then the change △Q of the charge of the capacitor is △Q=SHN L(a)e / (H/r)= SN L(a)d = (C - $C_0$)V, so the EDM of an Cs atom is **d = (C - $C_0$)V/ SN L(a).**

⑥If Cs atom has a large EDM, why has not been observed in other experiments? This is an interesting question. In Eq.(4) let the function f (a)= L(a)/a and from f ″(a)=0, we work out the knee of the function L(a)/a at $a_k=1.9296813\approx 1.93$. Corresponding knee voltage $V_k=V_2 a_k/a_2 \approx 11.4v$, $\log V_k=1.06$ and knee field $E_k=V_k/H_2 \approx 1.5\times 10^3 v/m$. By contrast with the curve in Fig.6, it is clear that our polarization equation is valid. The third experiment showed that the saturation polarization of Cs vapor is easily observed. When the saturation polarization of Cs atoms occurred ($V\geqslant 560$ volt), nearly all Cs atoms (more than 98％) are lined up with the field, and $C\approx C_o$ ($C_o$ is the vacuum capacitance) or the electric susceptibility of the Cs vapor was approximately the same as the vacuum! So the saturation polarization curve demonstrate especially clearly that only under the very weak field ($E<E_k$ i.e. $E<1.5\times 10^3 V/m$), the large EDM of Cs atom can be observed. **Regrettably, nearly all scientists in this field disregard the very important problem.**

⑦ As a concrete example, let us treat the linear Stark shifts of the hydrogen (n=2) and Cs atom. Notice that the fine structure of the hydrogen (n=2) is only 0.33 cm$^{-1}$ for the Hα lines of the Balmer series, and the fine structure is difficult to observe (where $\lambda = 656.3$ nm, and the splitting is only 0.014 nm)[21]. The linear Stark shifts of the energy levels is proportional to the field strength: $\triangle W= d.E=3ea_o E=1.59\times 10^{-8}$ E e.cm. When $E=10^5$ V/cm, $\triangle W=1.59\times 10^{-3}$ eV, this corresponds to a wavenumber of 12.8 cm$^{-1}$. So the linear Stark shifts is $\triangle\lambda=\triangle W\lambda^2/hc = 12.8\times(656.3\times 10^{-7})^2 =0.55$ nm. It is so large, in fact, that the linear Stark shift of the hydrogen(n=2) is easily observed [21]. Because the most field strength for Cs atoms is $E_{max}=7.4\times 10^4$ V/m, if Cs atom has a large EDM $d=3.5ea_o=1.86\times 10^{-8}$ e.cm, and the most splitting of the energy levels of Cs atoms $\triangle W_{max}= d.E_{max}= 3.5ea_o E_{max} =1.38\times 10^{-5}$ eV. This corresponds to a wavenumber of $11.1\times 10^{-2}$ cm$^{-1}$. On the other hand, the observed values for a line pair of the first primary series of Cs atom ($Z=55$, $n=6$) are $\lambda_1=894.35$ nm and $\lambda_2=852.11$ nm[12]. The magnitude of the linear Stark shift of Cs atoms is about $\triangle\lambda = \triangle W(\lambda_1+\lambda_2)^2/4hc = 0.0085$nm. **It is so small, in fact, that a direct observation of the linear Stark shifts of Cs atom is not possible.**

The striking feature of our experiments is the data are reliable and the measuring process can be easily repeated in any laboratory because the details of the experiment are described in the article. A detail explanation of how fill with Cs vapor at a fixed density in a vacuum environment as indicated in Fig.7. Our experimental apparatus are still kept, we welcome anyone who is interested in the experiments to visit and examine it. Experiments to search for the EDM of the ground state of Rubidium (Rb) atoms have been carried out. Similar results have been obtained and will be reported hereafter.

**Acnowledgement**   The authors thank to our colleagues Ri-Zhang Hu, Zhao Tang , Rui-Hua Zhou, Shao-Wei Peng, Yong Chen, Xue-ming Yi, Xiao-ming Wang, Xing Huang, and Engineer Jia You for their help in the work.

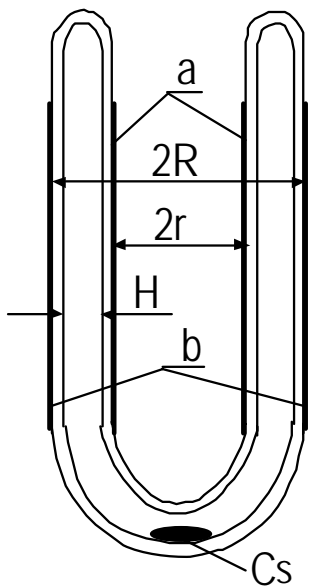
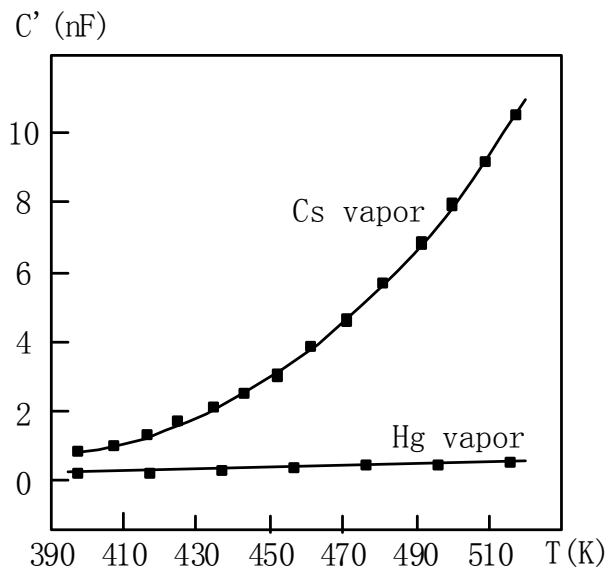

**Fig.1** This is the longitudinal section of the first experimental apparatus. Silver layers **a** and **b** build up a cylindrical glass capacitor. (not to scale).

**Fig.2** Two curves showed that the relationship between the capacitance C' of Cs or Hg vapor and the density N respectively, where $1nF=10^3$ pF.

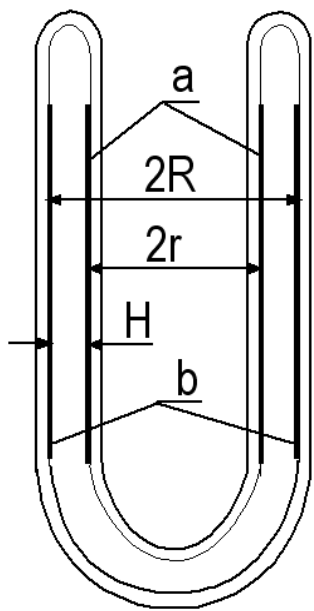
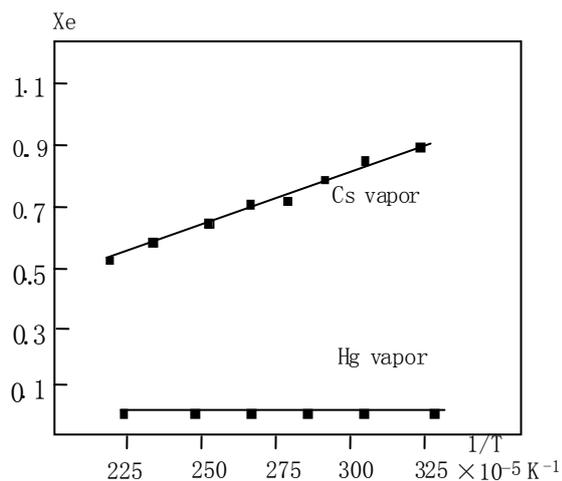

**Fig.3** This is the longitudinal section of another two experimental apparatus. Two round stainless steel tubes **a** and **b** build up a glass capacitor (not to scale).

**Fig.4** The temperature (T) dependence of the susceptibility ($x_e$) of Cs or Hg vapor. The slope of Cs vapor is B≈320(k) but the slope of Hg vapor is nearly zero, B≈0.0(k).

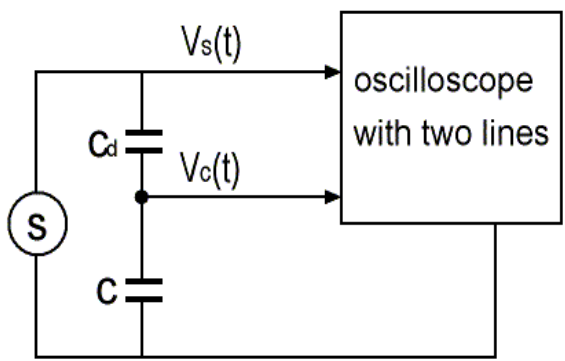
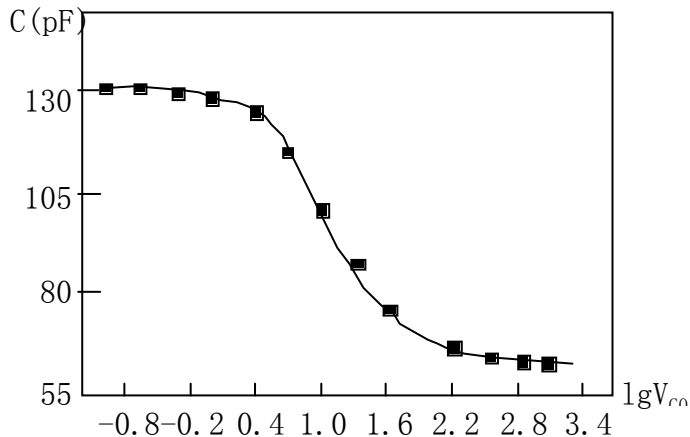

**Fig.5** The diagram shows the experimental method, in which C is capacitor filled with Cs vapor to be measured and $C_d$ is a standard one, where $V_s(t) = V_{so} \cos \omega t$ and $V_c(t) = V_{co} \cos \omega t$.

**Fig.6** The experimental curve shows that the saturation polarization effect of the Cs vapor is obvious when $E \geq 7.4 \times 10^4$ v/m.

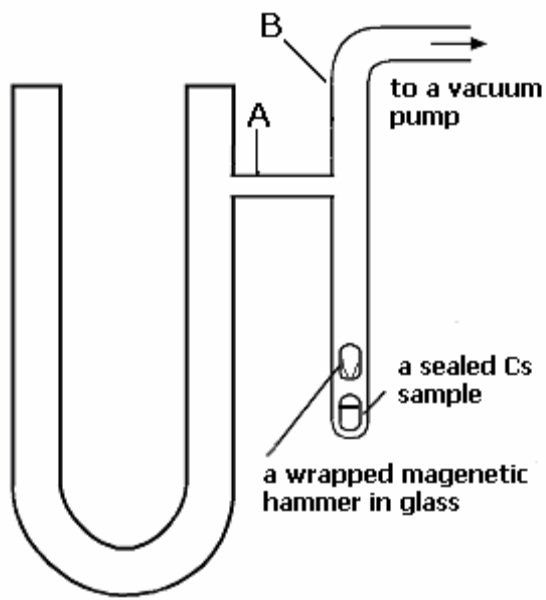

**Fig.7** The schematic diagram illustrates the major procedures. The experimental apparatus on the left is a closed glass containers resembling a Dewar bottle flask in shape. At first let the container and a vacuum pump linked by a glass tube. A sealed Cs sample in a small bottle with a hook-like mouth and a wrapped small magnetic hammer in glass are put together in the glass tube. When the Dewar flask is empty, it is pumped to vacuum pressure $P \leq 10^{-6}$ Pa for twenty hours in temperature range 20℃～220℃. In vacuum environment the small magnetic hammer is raised by a magnet which resembles letter "U" in shape outside the glass tube. When the hammer is released suddenly, it breaks the hook-like mouth of the bottle that sealed Cs sample, then the glass tube is sealed at the position of B point. Next step, the apparatus is again heated up to 220℃. When Cs material in liquid state sink to the bottom of the tube and the Dewar flask was filled with volatile Cs vapor, the glass tube was sealed once again at the position of A point. Thus we obtain a cylindrical capacitor filled with Cs vapor at a fixed density (not to scale).